\global\def\draftcontrol{0}

   \def\versionno{Heisenberg in Quivers}

\catcode`\@=11

\expandafter\ifx\csname draftcontrol\endcsname\relax\global\def\draftcontrol{0}
\fi

{\count255=\time\divide\count255 by 60
\xdef\hourmin{\number\count255}
\multiply\count255 by-60\advance\count255 by\time
\xdef\hourmin{\hourmin:\ifnum\count255<10 0\fi\the\count255}}
\def\draftdate{\number\month/\number\day/\number\year\ \ \ \hourmin }

\newcommand\makepapertitle{\par
  \begingroup
    \renewcommand\thefootnote{\@fnsymbol\c@footnote}%
    \def\@makefnmark{\rlap{\@textsuperscript{\normalfont\@thefnmark}}}%
    \long\def\@makefntext##1{\parindent 1em\noindent
            \hb@xt@1.8em{%
                \hss\@textsuperscript{\normalfont\@thefnmark}}##1}%
     \newpage
     \global\@topnum\z@   
     \@makepapertitle
     \thispagestyle{empty}\@thanks
  \endgroup
  \setcounter{footnote}{0}%
  \global\let\thanks\relax
  \global\let\makepapertitle\relax
  \global\let\@makepapertitle\relax
  \global\let\@thanks\@empty
  \global\let\@author\@empty
  \global\let\@date\@empty
  \global\let\@title\@empty
  \global\let\title\relax
  \global\let\author\relax
  \global\let\date\relax
  \global\let\and\relax
  \def\version{\let\version\@version\@gobble}
}
\def\@makepapertitle{%
  \newpage
   \ifnum\draftcontrol=1 {}
   \version\versionno
   \vskip 3em%
   \else
   \hfill\hbox to 3cm {\parbox{4cm}{\@pubnum}\hss}%
   \vskip 3em%
   \fi
   \begin{center}%
   \let \footnote \thanks
     {\LARGE {\@title}}%
     \vskip 1.5em%
     {\normalsize
       \lineskip .5em%
       \begin{tabular}[t]{c}%
         \@author
       \end{tabular}\par}%
     \vskip 1.5em%
     {\@bstract}%
     \end{center}%
     \vskip 1.5em
     \@date%
   \par
}

\gdef\@pubnum{}
\def\pubnum#1{%
  \gdef\@pubnum{#1}}

\gdef\@bstract{}
\def\Abstract#1{%
  \gdef\@bstract{%
   \parbox{\textwidth-0pc}{%
   \centerline{\bf Abstract}\penalty1000%
\noindent
\renewcommand\baselinestretch{1.0}%
{#1}}}
}

\def\ps@paper{\let\@mkboth\@gobbletwo%
     \ifnum\draftcontrol=1
        \def\@oddfoot{\hbox to \textwidth{\tiny \versionno \hfil\tiny\draftdate}%
        \hskip -\textwidth \hbox to \textwidth{\hfil\rm\thepage\hfil}}%
     \else\def\@oddfoot{\hbox to \textwidth{\hfil\rm\thepage\hfil}}
     \fi
     \let\@evenfoot\@oddfoot
}

\def\@version#1{\ifnum\draftcontrol=1
\typeout{}\typeout{#1}\typeout{}
\vskip3mm\centerline{\hbox{\fbox{\normalsize{\tt DRAFT -- #1 -- }
                   {\draftdate}}}}\vskip3mm
\fi}
\let\version\@version
\long\def\eqlabel#1{\ifnum\draftcontrol=1
                    \tag@false  
                    \tag*{(\theequation) \hbox to -0.2cm{\hspace{0cm}\small{#1}\hss}}
                    \refstepcounter{equation}
                    \edef\@currentlabel{\theequation}
                    \ltx@label{#1}          
                    \else
                    \label{#1}
                    \fi
                    }
\let\st@bibitem\@bibitem
\let\st@lbibitem\@lbibitem
\ifnum\draftcontrol=1
  \def\@bibitem#1{%
    \st@bibitem{#1}\a@@label{#1}\ignorespaces}
  \def\@lbibitem[#1]#2{%
    \st@lbibitem[#1]{#2}\a@@label{#2}\ignorespaces}
  \def\a@@label#1{%
    \gdef\a@lab{\smash{\normalfont\small#1}}
    \ifvmode
      \if@inlabel
        \global\setbox\@labels\hbox{%
          \llap{\a@lab\let\a@lab\relax
                \kern\@totalleftmargin\kern\marginparsep}%
          \box\@labels}%
      \fi
    \fi}
\fi

\documentclass[12pt,letterpaper]{article}

\usepackage{amsmath,amssymb,array,calc,rotating,epsfig,psfrag}
\usepackage[nosort]{cite}

\ifnum\draftcontrol=1
\fi

\renewcommand\baselinestretch{1.25}
\renewcommand\section{\@startsection {section}{1}{\z@}%
                                   {-3.5ex \@plus -1ex \@minus -.2ex}%
                                   {2.3ex \@plus.2ex}%
                                   {\normalfont\large\bfseries}}
\renewcommand\subsection{\@startsection{subsection}{2}{\z@}%
                                   {-3.25ex\@plus -1ex \@minus -.2ex}%
                                   {1.5ex \@plus .2ex}%
                                   {\normalfont\normalsize\bfseries}}
\renewcommand\subsubsection{\@startsection{subsubsection}{3}{\z@}%
                                   {-3.25ex\@plus -1ex \@minus -.2ex}%
                                   {1.5ex \@plus .2ex}%
                                   {\normalfont\normalsize\it}}
\renewcommand\paragraph{\@startsection{paragraph}{4}{\z@}%
                                   {-3.25ex\@plus -1ex \@minus -.2ex}%
                                   {1.5ex \@plus .2ex}%
                                   {\normalfont\normalsize\bf}}

\def\revise#1       {\raisebox{-0em}{\rule{3pt}{1em}}%
                     \marginpar{\raisebox{.5em}{\vrule width3pt\
                     \vrule width0pt height 0pt depth0.5em
                     \hbox to 0cm{\hspace{0cm}{%
                     \parbox[t]{4em}{\raggedright\footnotesize{#1}}}\hss}}}}

\def\complex      {{\mathbb C}}

\def\zet          {{\mathbb Z}}

\def\del          {\partial}

\def\tr           {\mathop{\rm Tr}}

\def\half{{\frac12}}

\def\sqr#1#2{{\vcenter{\vbox{\hrule height.#2pt
 \hbox{\vrule width.#2pt height#1pt \kern#1pt
 \vrule width.#2pt}\hrule height.#2pt}}}}

\def\a{\alpha}
\def\b{\beta}
\def\r{\rho}

\def\m{\mu}
\def\g{\gamma}

\def\n{\nu}
\def\bn{\bar{\nu}}
\def\bm{\bar{\mu}}

\newcommand{\Tr}{{\rm Tr\,}}

\catcode`\@=12

\usepackage{color}

\begin{document}




\newcommand{\be}{\begin{equation}}
\newcommand{\ee}{\end{equation}}
\newcommand{\beq}{\begin{equation}}
\newcommand{\eeq}{\end{equation}}
\newcommand{\ba}{\begin{eqnarray}}
\newcommand{\ea}{\end{eqnarray}}
\newcommand{\nn}{\nonumber}

\def\vol{\bf vol}
\def\Vol{\bf Vol}
\def\del{{\partial}}
\def\vev#1{\left\langle #1 \right\rangle}
\def\cn{{\cal N}}
\def\co{{\cal O}}
\def\IC{{\mathbb C}}
\def\IR{{\mathbb R}}
\def\IZ{{\mathbb Z}}
\def\RP{{\bf RP}}
\def\CP{{\bf CP}}
\def\Poincare{{Poincar\'e }}
\def\tr{{\rm tr}}
\def\tp{{\tilde \Phi}}
\def\Y{{\bf Y}}
\def\te{\theta}
\def\bX{\bf{X}}

\def\TL{\hfil$\displaystyle{##}$}
\def\TR{$\displaystyle{{}##}$\hfil}
\def\TC{\hfil$\displaystyle{##}$\hfil}
\def\TT{\hbox{##}}
\def\HLINE{\noalign{\vskip1\jot}\hline\noalign{\vskip1\jot}} 
\def\seqalign#1#2{\vcenter{\openup1\jot
  \halign{\strut #1\cr #2 \cr}}}
\def\lbldef#1#2{\expandafter\gdef\csname #1\endcsname {#2}}
\def\eqn#1#2{\lbldef{#1}{(\ref{#1})}%
\begin{equation} #2 \label{#1} \end{equation}}
\def\eqalign#1{\vcenter{\openup1\jot
    \halign{\strut\span\TL & \span\TR\cr #1 \cr
   }}}
\def\eno#1{(\ref{#1})}
\def\href#1#2{#2}
\def\half{{1 \over 2}}

\def\ads{{\it AdS}}
\def\adsp{{\it AdS}$_{p+2}$}
\def\cft{{\it CFT}}

\newcommand{\ber}{\begin{eqnarray}}
\newcommand{\eer}{\end{eqnarray}}

\newcommand{\bea}{\begin{eqnarray}}
\newcommand{\eea}{\end{eqnarray}}

\newcommand{\beqar}{\begin{eqnarray}}
\newcommand{\cN}{{\cal N}}
\newcommand{\cO}{{\cal O}}
\newcommand{\cA}{{\cal A}}
\newcommand{\cT}{{\cal T}}
\newcommand{\cF}{{\cal F}}
\newcommand{\cC}{{\cal C}}
\newcommand{\cR}{{\cal R}}
\newcommand{\cW}{{\cal W}}
\newcommand{\eeqar}{\end{eqnarray}}
\newcommand{\lm}{\lambda}\newcommand{\Lm}{\Lambda}
\newcommand{\eps}{\epsilon}


\newcommand{\nonu}{\nonumber}
\newcommand{\oh}{\displaystyle{\frac{1}{2}}}
\newcommand{\dsl}
  {\kern.06em\hbox{\raise.15ex\hbox{$/$}\kern-.56em\hbox{$\partial$}}}
\newcommand{\as}{\not\!\! A}
\newcommand{\ps}{\not\! p}
\newcommand{\ks}{\not\! k}
\newcommand{\D}{{\cal{D}}}
\newcommand{\dv}{d^2x}
\newcommand{\Z}{{\cal Z}}
\newcommand{\N}{{\cal N}}
\newcommand{\Dsl}{\not\!\! D}
\newcommand{\Bsl}{\not\!\! B}
\newcommand{\Psl}{\not\!\! P}
\newcommand{\eeqarr}{\end{eqnarray}}
\newcommand{\ZZ}{{\rm \kern 0.275em Z \kern -0.92em Z}\;}

\def\s{\sigma}
\def\a{\alpha}
\def\b{\beta}
\def\r{\rho}
\def\d{\delta}
\def\g{\gamma}
\def\G{\Gamma}
\def\ep{\epsilon}
\makeatletter \@addtoreset{equation}{section} \makeatother
\renewcommand{\theequation}{\thesection.\arabic{equation}}

\def\be{\begin{equation}}
\def\ee{\end{equation}}
\def\bea{\begin{eqnarray}}
\def\eea{\end{eqnarray}}
\def\m{\mu}
\def\n{\nu}
\def\g{\gamma}
\def\p{\phi}
\def\L{\Lambda}
\def \W{{\cal W}}
\def\bn{\bar{\nu}}
\def\bm{\bar{\mu}}
\def\bw{\bar{w}}
\def\ba{\bar{\alpha}}
\def\bb{\bar{\beta}}

\begin{titlepage}

\version\versionno

\leftline{\tt hep-th/0602094}

\vskip -.8cm

\rightline{\small{\tt MCTP-06-01}}
\rightline{\small{\tt NSF-KITP-06-07}}

\vskip 1.7 cm

\centerline{\bf \Large Finite Heisenberg Groups in Quiver Gauge Theories}

\vskip 1cm
\vskip 1cm
{\large }
\vskip 1cm

\centerline{\large Benjamin A. Burrington, James T. Liu, and Leopoldo A. Pando
Zayas }

\vskip 1cm
\centerline{\it Michigan Center for Theoretical
Physics}
\centerline{ \it Randall Laboratory of Physics, The University of
Michigan}
\centerline{\it Ann Arbor, MI 48109-1120}

\vspace{1cm}

\begin{abstract}
We show by direct construction that a large class of quiver gauge
theories admits actions of finite Heisenberg groups. We
consider various quiver gauge theories that arise as AdS/CFT duals
of orbifolds of $\mathbb{C}^3$, the conifold and its
orbifolds and some orbifolds of the cone over $Y^{p,q}$. Matching the
gauge theory analysis  with string theory on the corresponding spaces
implies that the operators counting wrapped branes do not commute in
the  presence of flux.

\end{abstract}



\end{titlepage}


%

\section{Introduction}
The AdS/CFT has generally been used to obtain information about
strongly coupled gauged theories using the weakly coupled supergravity
side \cite{agmoo}. Given that our technology for understanding string
theory in backgrounds with Ramond-Ramond fluxes is still inadequate it
could be advantageous to use gauge theories to understand fundamental
properties of string theory such as the nature of D branes. In fact,
in a very interesting paper Gukov, Rangamani and Witten did just that
\cite{grw}. They matched operators on the field theory defined as a
$\mathbb{Z}_3$ orbifold of ${\cal N}=4$ super Yang Mills to wrapped
branes in type IIB string theory on $AdS_5\times S^5/\mathbb{Z}_3$
with five-form flux, therefore uncovering a very interesting
noncommutative structure. They found a finite Heisenberg group realized by
discrete transformations in the gauge theory. These transformations
can subsequently be interpreted as operators counting various wrapped
branes in the string theory.

The idea that D-brane charge in the presence of flux might not be a
commutative quantity has far reaching  implications for our
understanding of D-branes. It has recently been argued by D. Belov,
G. Moore and others \cite{kitp} that this is, in fact, the general
situation whenever the space has homology groups with nontrivial
torsion classes.  An interesting review of the mathematical background
can be found in \cite{freed}.

If the structure uncovered in \cite{grw} is generic for D-branes in
backgrounds with RR fluxes then we should be able to display it in a
more general setting. With this motivation in mind we turned to
generalizations of the construction of
\cite{grw}. We show by direct construction that a large class of quiver gauge
theories admits an action of the  Heisenberg group. In particular we
consider various quiver gauge theories that arise as duals in the
AdS/CFT sense  of orbifolds of $\mathbb{C}^3$, the conifold and its
orbifolds and some orbifolds of the cone over $Y^{p,q}$.

Our main result can be formulated as follows. For a large class of
quiver gauge theories with gauge group $SU(N)^p$, there is a set of
discrete transformations $A, B$ and $C$ satisfying
\be
A^{q}=B^{q}=C^{q}=1, \qquad AB=BAC.
\ee
where $q$ is some integer
number which depends on the particular structure of the quiver.  These
transformations satisfy three important properties: (i) leave the
superpotential invariant, (ii) satisfy the anomaly cancelation for all
$SU(N)$ gauge groups, and (iii) the above group relations are true up to
elements in the center of the gauge group $SU(N)^p$, that is, up to gauge transformations.

This structure can be interpreted in terms of  D branes in the dual
string theory. In this case the operators $A,  B$ and $C$ count the
number of wrapped fundamental strings, D-strings and D3 branes
respectively.  The matching is impressive, in particular the number
$q$ above is related to torsion classes in the third homology group
$H_3(X,\mathbb{Z})=\mathbb{Z}_q$, where $X$ is the horizon manifold of
the space dual to the quiver gauge theory.

The organization of this note is as follows. In section two we present
the guiding principles in the search for the set of symmetries
realizing finite Heisenberg groups in some quiver gauge theories. After the
general setup we consider orbifolds of ${\cal N}=4$ SYM, these
theories can be understood in the AdS/CFT sense as orbifolds of
$\mathbb{C}^3$; we also discuss the conifold and its orbifolds and
some orbifolds of the cone over $Y^{p,q}$. Section three contains
comments on the D-brane interpretation. We conclude in section four.

\section{Finite Heisenberg groups in quiver gauge theories}
In this section we discuss the set up of the general construction. We want to
display how the constrains of classical invariance of the
superpotential
and anomaly cancelation determine the symmetries.

As mentioned in the introduction, we seek discrete transformations $A, B$ and $C$ of the gauge
theory. First, these transformations are discrete and cyclical:
\be
A^{q}=B^{q}=C^{q}=1,
\ee
where $q$ is an integer that depends on the concrete theory.
More interestingly, these transformation satisfy a finite Heisenberg
group relation
\be
AB=BAC,
\ee
and the element $C$ commutes with $A$ and $B$.

We take a constructive approach. As a first step
we  consider a special class of quiver diagrams that
admit a shift symmetry. This means that there is an obvious map
of fields to fields and gauge groups to gauge groups.  In the case of a $\mathbb{Z}_q$ orbifold
of ${\cal N}=4$ SYM one has $q$ gauge groups. The $A$ symmetry is a cyclic permutation of the gauge groups.
There is an alternative way to view the appearance of the $A$ transformation from the dual string theory side.
For example, the correspondent dual is described by
$N$ D3 branes near a $\complex^3/\zet_q$
singularity.  In the string theory setup there is a natural shift symmetry.  The shift symmetry is just the
action of moving the stack of branes onto their image  (and so
this symmetry is a $\zet_q$).
 The operators $B$ and $C$ are then realized as
``rephasing'' symmetries which multiply the superfields, component by
component, by a constant phase.

To begin discussion of these symmetries, we set our labeling
conventions.  We will label vertices with numbers.  Superfields will
then be labeled with an upper case letter carrying a subscript that
denotes which vertex it originates from (or goes to, depending on the case).
If there is a global
symmetry under which a set of fields transforms, then this index will
be labeled by a superscript.  The phase that we shift a field by will
be denoted by lowercase letter and the same subscript that labels the field.  For
example, in a quiver diagram with a global $SU(2)$ symmetry, the
fields will be labeled

\begin{figure}[h]
\centering
\includegraphics[width=.35\textwidth]{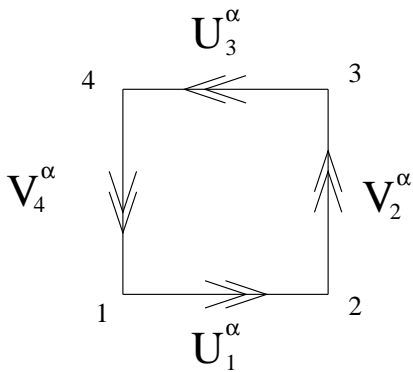}
\caption{Example of a quiver diagram with global symmetries.}
\label{exam}
\end{figure}

where the $U^\alpha_n$ and $V^\alpha_n$ transform as doublets under the
$SU(2)$.  To preserve a global symmetry we must scale all fields in a
representation by the same phase, e.g. $U^\alpha_1 \mapsto u_1
U^\alpha_1$.  We again emphasize that this rephasing of the fields to
act on the component fields  separately, and so we are not combining
this with an R symmetry associated with the Grassmann variables.

For this to be a classical symmetry, we require invariance of the
superpotential.  In general, one can read the superpotential terms by
going in loops in a diagram, making sure that the fields can be
combined into invariant terms under any global symmetry.  We
will only consider loops that include 3 and 4 fields.  For example, in
figure (\ref{exam}) there is only one superpotential term
$\propto \epsilon_{\alpha \beta}\epsilon_{{\alpha'} {\beta'}}
\Tr(U^\alpha_1 V^{\alpha'}_2 U^\beta_3 V^{\beta'}_4)$.  Invariance of the superpotential implies
that we must demand that classically $u_1 v_2 u_3 v_4=1$.  This
generalizes easily to other quivers. For a monomial term in a
superpotential one replaces the fields by the associated scalings and
requires this product to be 1:
\be
{\mathcal W}=\cdots+g\Tr\left(U_1 U_2 \cdots U_k \right)+\cdots
\longrightarrow u_1u_2\cdots u_k=1,\;\;\cdots.
\ee
Now we look for those symmetries that are not broken by quantum
effects.  The problematic terms in the path integral measure are the
fermions.  The basic point is that in instanton backgrounds, the Dirac
operator has fermion zero modes while the conjugate Dirac operator
does not.  This leads to an
asymmetry in the path integral measure, and so the above scalings will,
in general, be anomalous.  The number of fermion zero modes in a background with
instanton number $J$ is  $2T({\bf r})\times J$.  One can
see this dependance on $T({\bf r})$ coming from the
$(j^\mu_{phase},A_{SU(N)},A_{SU(N)})$ triangle anomaly.  The most
restrictive case for us is when $J=1$ because this will raise the
scaling to the smallest power.  Let us take an example where the
$SU(N)$ associated with vertex $1$ has instanton number 1.  To
consider the anomaly, we consider the part of the diagram around this
$SU(N)$ :

\begin{figure}[h]
\centering
\includegraphics[width=.35\textwidth]{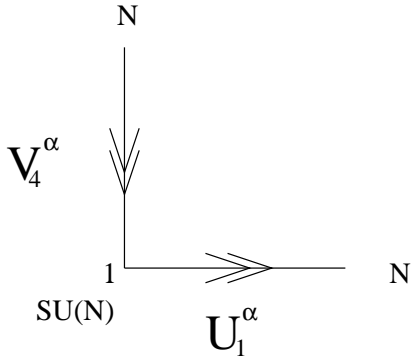}
\caption{Anomaly contribution for a $SU(N)$ factor in the working example.}
\end{figure}

Note that the end of the arrows have an $N$ fold degeneracy due to the
$SU(N)$ gauge groups not under consideration.  Therefore, there are
$N\times2 \times 2 T({\bf r})$ fermion zero modes associated with each
of the fields $U_1$ and $V_4$.  These are (anti) fundamentals, and so
$T({\bf r})=1/2$.  We conclude that the measure transforms unless
$(v_4^2 u_1^2)^{N}=1$.  One may repeat this calculation for the other
vertices, and one finds that $(u_i^2 v_{i\pm1}^2)^N$ for $i=1,3$ and
the subscripts are to be read mod 4.  For a general diagram one considers
one gauge group at a time, and considers the fields that couple to
this gauge group.  Then, one counts the multiplicity of the fields by
the number of arrows times the rank of the gauge group at the other
end of the arrow.  The condition that this is a symmetry is that
$\prod a_i^{(mult.)_i}=1$.  For example, a vertex in a quiver diagram

\begin{figure}[h]
\centering
\includegraphics[width=.5\textwidth]{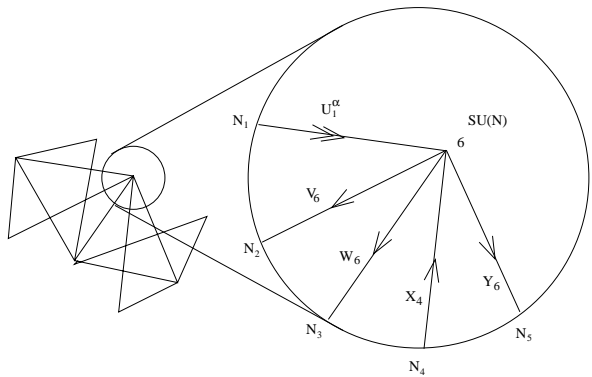}
\caption{Anomaly contribution for a $SU(N)$ factor in the general case.}
\end{figure}

\be
u_1^{2N_1}v_6^{N_2}w_6^{N_3}x_4^{N_4}y_6^{N_5}.
\ee

We now turn to a particular case of the above symmetries that are
related to the gauge symmetry.  We will consider only quiver theories
that have an $SU(N)$ at every vertex.  For quiver diagrams with $n$ vertices the gauge group
is then  $SU(N)^n$ with center $(\zet_N)^n$.  The center of the
gauge group corresponds to scaling each field by an $N^{\mbox{th}}$
root of unity, but in such a way as to leave the superpotential
invariant.  These then fall into the category just described.  The
gauge symmetry is a redundancy, and so we identify the scalings
described in the last paragraph up to elements of the center of the
gauge group.

\subsection{Orbifolds of $\mathbb{C}^3$}
The gauge theories obtained by orbifolding ${\cal N}=4$ where
discussed in the context  of the AdS/CFT by \cite{ks,lnv}. The
techniques were essentially developed by Douglas and Moore in
\cite{dm}.

Let us first discuss the action of the orbifold on $\mathbb{C}^3$. Let
 the complex coordinate on  $\mathbb{C}^3$ be $z_1, z_2$ and $z_3$,
 the orbifold of $\mathbb{Z}_p$ acts as
\be
(z_1,z_2,z_3)\mapsto (\omega z_1, \omega z_2, \omega^{-2} z_3),
\ee
where $\omega$ is a $p$-th root of unity. It is important that this
transformation preserves the natural  holomorphic $(3,0)$-form the
determines the Calabi-Yau structure, that is, it leaves $dz_1\wedge dz_2 \wedge dz_3$ invariant.

Let us consider the case $\mathbb{Z}_5$ for concreteness but the techniques generalize directly.
The general idea for orbifolding a theory consist on working in its covering space.
In the case of $SU(N)$ gauge group therefore consider a theory with gauge group $SU(N)^5$.

The action of the orbifold treats coordinates $z_1$ and $z_2$ identically, in the quiver diagram this
implies that we have a doublet under $SU(2)$ which we denote by $U_i^{\alpha}$, where $\a$ is the $SU(2)$ index.

The superpotential terms can be obtained by traveling in loops in the quiver diagram.

\be
W=\epsilon_{\a\b}U_1^\a U_2^\b Z_1-\epsilon_{\a\b}U_2^\a U_3^\b Z_2 + \epsilon_{\a\b}U_3^\a U_4^\b Z_3
-\epsilon_{\a\b}U_4^\a U_5^\b Z_4+\epsilon_{\a\b}U_5^\a U_1^\b Z_5.
\ee

\begin{center}
\includegraphics[width=.5\textwidth]{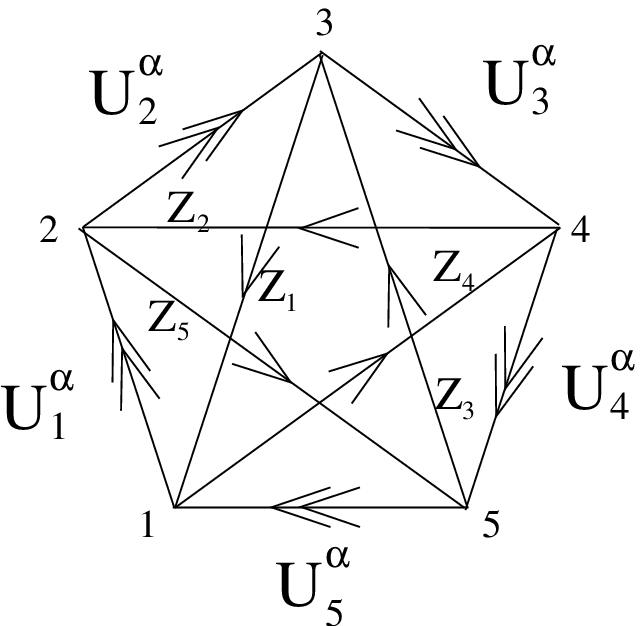}
\end{center}

The $A$ symmetry is easy to spot in this case:
\be
A:
\left(
\begin{array}{c}
U_i^{\a}\\
Z_i\\
\end{array}
\right)
\mapsto
\left(
\begin{array}{c}
U_{i+1}^\a\\
Z_{i+1}\\
\end{array}
\right).
\ee
This operation obviously has the property $A^5=1$.

We now proceed to look for rephasing of the fields that leave the
theory invariant. The general transformation, as explained  at the
beginning of this section involves: $U_i \mapsto u_i U_i$ and $Z_i
\mapsto z_i Z_i$, where $u_i$ and $z_i$ are some root  of unity.
Invariance of the superpotential implies an interesting factorization
of the problem. For example, for the first  term in the superpotential
(the loop starting in the lower left corner) we have that
\be
z_1 u_1 u_2=1.
\ee
We view this, and similar equalities as determining the
scaling of the $Z$ fields once the scalings of the $U$ fields is
known. The anomaly condition for the lower left corner gives:
\be
\left(u_5^2z_4 z_1 u_1^2\right)^N=1,
\ee
we can further eliminate the $z's$ to obtain expressions of the form:
\be
\left(\frac{u_1 u_5}{u_2 u_4}\right)^N=1.
\ee
Similar expressions are obtained with a cyclical reordering of
the indices. Note that an interesting pattern arises:
product of two consecutive phases equals the product of the phase
before and after the sequence. We could now look for solutions to the
above equations and find the $B$ and $C$ transformation. Here we will
follow a different path that automatically eliminates some issues with
transformations in the center of the gauge group.

To identify the scaling symmetries, we find it convenient to consider certain members
of the center of the gauge group.  To do so, we associate an integer $n_i$ with each vertex.
This integer tells which member of the center of that gauge group we are rephasing by:
$e^{\frac{2 \pi i n_i}{N}}{\mathbb I}_{N\times N}$.  This then gives us a prescription
for how to rephase the fields in a gauge invariant way.  We consider the two scalings
given by
\begin{center}
\includegraphics[width=.5\textwidth]{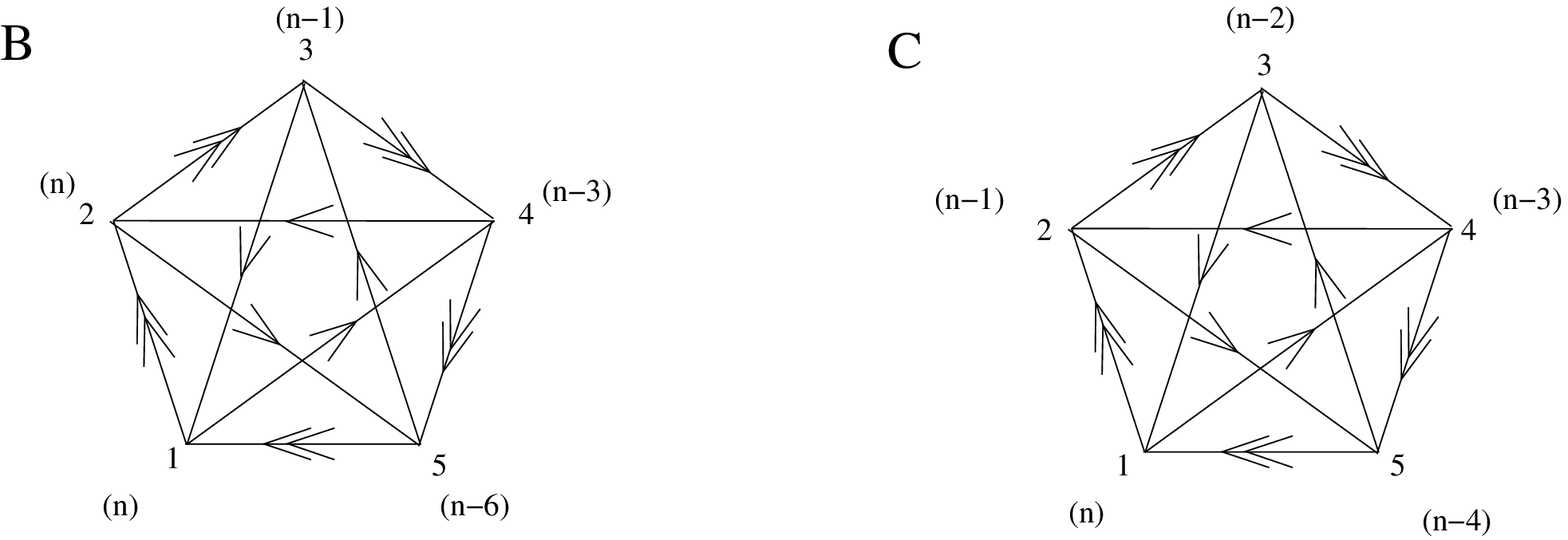}
\end{center}
which translate to
\be
B:
\begin{array}{ccl}
(U_1,U_2,U_3,U_4,U_5)&\mapsto &(U_1, \g U_2, \g^2 U_3, \g^3 U_4, \g^{-6} U_5) \\
(Z_1,Z_2,Z_3,Z_4,Z_5)&\mapsto & (\g^{-1} Z_1,\g^{-3} Z_2,\g^{-5}Z_3, \g^3 Z_4,\g^6 Z_5)
\end{array}
\ee
\be
C:
\begin{array}{ccl}
(U_1,U_2,U_3,U_4,U_5)&\mapsto &(\g U_1, \g U_2, \g U_3, \g U_4, \g^{-4} U_5) \\
(Z_1,Z_2,Z_3,Z_4,Z_5)&\mapsto & (\g^{-2} Z_1,\g^{-2} Z_2,\g^{-2} Z_3, \g^3 Z_4,\g^3 Z_5)
\end{array}
\ee
for the fields.  Note that the above transformations were picked to have $B$ as close to a ``clock''
symmetry, and $C$ was designed ``undo'' a shift of the above fields.  The structure of the
$Z$ field transformations come as a consequence of the $U$ fields.
Finally, the above transformations are a redundancy because $\g^N=1$, i.e. is in the center of the gauge
group.  However, to find non trivial elements, we wish to promote this action so that it is not in the
center of the gauge
group.  We consider instead
\be
B:
\begin{array}{ccl}
(U_1,U_2,U_3,U_4,U_5)&\mapsto &(U_1, \omega U_2, \omega^2 U_3, \omega^3 U_4, \omega^{-6} U_5) \\
(Z_1,Z_2,Z_3,Z_4,Z_5)&\mapsto & (\omega^{-1} Z_1,\omega^{-3} Z_2,\omega^{-5}Z_3, \omega^3 Z_4,\omega^6 Z_5)
\end{array}
\ee
\be
C:
\begin{array}{ccl}
(U_1,U_2,U_3,U_4,U_5)&\mapsto &(\omega U_1, \omega U_2, \omega U_3, \omega U_4, \omega^{-4} U_5) \\
(Z_1,Z_2,Z_3,Z_4,Z_5)&\mapsto & (\omega^{-2} Z_1,\omega^{-2} Z_2,\omega^{-2} Z_3, \omega^3 Z_4,\omega^3 Z_5)
\end{array}
\ee
Considering the anomaly condition at the vertices, we conclude that the above
transformations are anomaly free if $\omega^{5N}=1$.
For example, considering node $1$, the fermion measure transforms as
\be
\omega^{2(-4)N}\omega^{2N}\omega^{-2N}\omega^{3N}=\omega^{-5N}
\ee
under transformation $C$, and
\be
\omega^{2(-6)N}\omega^{3N}\omega^{-N}=\omega^{-10N}
\ee
under $B$.  We we demand these to be 1.  This gives us immediately that
we take $B^5=C^5=1$ because these are in the center of the gauge group.

One may also check that in the $U$ sector
\be
A_U C_U=C_U A_U\times \begin{pmatrix}
\omega^5 &0&0&0&0 \\
0&1&0&0&0 \\
0&0&1&0&0 \\
0&0&0&1&0 \\
0&0&0&0&\omega^{-5}
\end{pmatrix}\equiv C_U A_U \times M_U
\ee
and that in the Z sector
\be
A_Z C_Z=C_Z A_Z\times \begin{pmatrix}
\omega^{-5} &0&0&0&0 \\
0&1&0&0&0 \\
0&0&1&0&0 \\
0&0&0&\omega^{5}&0 \\
0&0&0&0&0
\end{pmatrix}\equiv C_Z A_Z \times M_Z.
\ee
The matrices above are in the center of the gauge group
with
\begin{center}
\includegraphics[width=.35\textwidth]{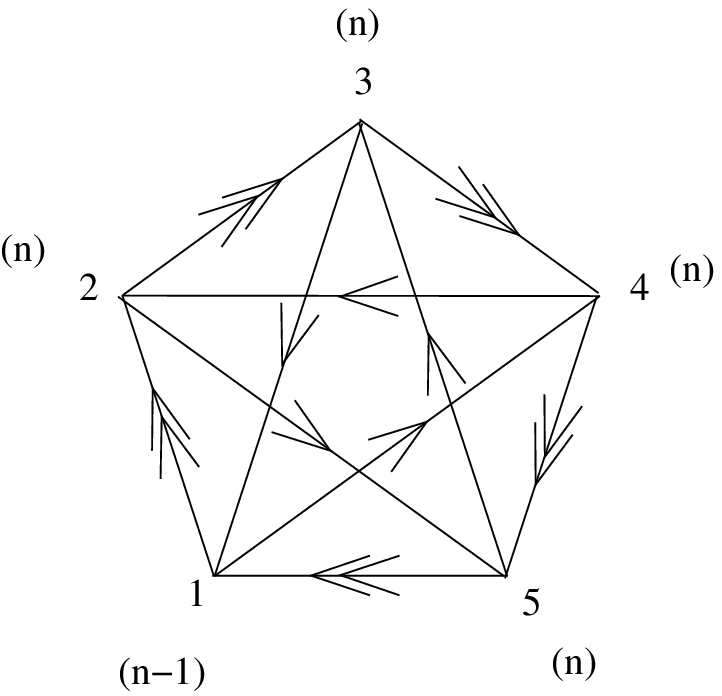}
\end{center}
and $\g=\omega^5$.  It is also easy to check that
\bea
A_U B_U&=&B_U A_U C_U \times M_U \nn \\
A_Z B_Z&=&B_Z A_Z C_Z \times M_Z.
\eea
The extra factors on the right are in fact part of the center of the gauge group as
before.  Therefore, we see that we have
\be
AB=BAC, \quad AC=CA,\quad BC=CB, \quad A^5=B^5=C^5=1.
\ee
where the equalities are to be read only up to the center of the gauge group.

\subsection{The conifold  and its Orbifolds}
The gauge theory of the worldvolume of $N$ D3 branes in the conifold
singularity was introduced by  Klebanov and Witten \cite{kw}. It
admits a holographic description in terms of IIB string theory on
$AdS_5\times T^{1,1}$. The theory has gauge group $SU(N)\times SU(N)$
and matter content described by two fields $U^\alpha$ and $V^\alpha$
which are doubles with respect to $SU(2)$ and transform in the
bifundamental representation.

\begin{center}
\begin{tabular}[h]{|l||c|c|}
\hline
         & $SU_1(N)$  &$SU_2(N)$  \\
\hline
\hline
$U_1^\alpha$ & $N$ & $\bar N$ \\
\hline
$V_2^\alpha $ & $\bar N$ & $ N$  \\
\hline
\end{tabular}
\end{center}

A simple way to represent this theory is its quiver diagram

\begin{center}
\includegraphics[width=.35\textwidth]{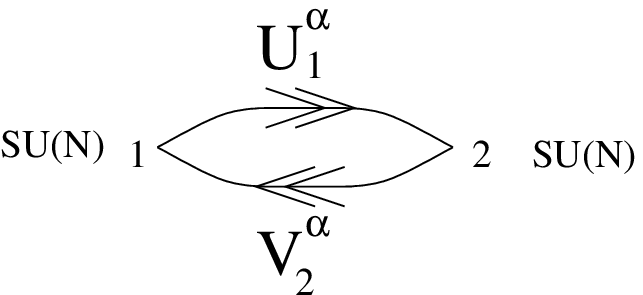}.
\end{center}

The superpotential of the theory is
\be
W= \epsilon_{\a\gamma}\epsilon_{\beta \delta} U^\a V^\b U^\gamma V^\delta.
\ee

A natural candidate for the $A$ transformation is $A:(U,V)\mapsto
(V,U)$.

We can start by considering arbitrary discrete transformation of the
form

\be
(U,V) \mapsto (u U, v V).
\ee

The condition that this transformation preserves the superpotential
implies that
\be
u^2 v^2=1.
\ee

The absence of anomalies implies that
\be
u^{2N} v^{2N}=1.
\ee

For $\omega$ a square root of unity we have.
\begin{eqnarray}
A&:&(U,V)\mapsto (V,U), \\ B&:& (U,V)\mapsto (U, \omega V), \\\ C&:&
(U,V) \mapsto (\omega U, \omega^{-1} V).
\end{eqnarray}

It is easy to show that $A, B$ and $C$ satisfy the relations of a
finite Heisenberg group
with four elements.

There is, however, a problem with our transformation. The
superpotential actually transforms  by a minus sign.  The fact that
the superpotential transforms by a minus sign under $(U,V) \mapsto (V,U)$ was already pointed out in
\cite{kw}. This transformation can, in principle, be accompanied by
and $R$-symmetry transformation $\theta\mapsto i\theta$ which
compensates. Thus we could conclude that the Heisenberg group of the
conifold is anomalous unless it can be accompanied by an R-symmetry
rotation. This is more like an accidental symmetry, we will comment on
this in the next section.

\subsubsection{$\mathbb{Z}_2$ orbifold of the conifold}
The gauge theory living in the worldvolume of $N$ D3 branes placed at
the singularity of the  $\mathbb{Z}_2$ orbifold can be obtained from
the previous construction. For the case of the $\mathbb{Z}_2$
orbifold we have that the superpotential is
\be
W = \epsilon_{\a\gamma}\epsilon_{\beta \delta} {\rm Tr} U_1^\a V_2^\beta U_3^\gamma V_4^\delta.
\ee

The transformation rules for the fields are
\begin{center}
\begin{tabular}[h]{|l||c|c|c|c|}
\hline
         & $SU_1(N)$  &$SU_2(N)$ & $SU_3(N)$  &$SU_4(N)$ \\
\hline
\hline
$U^\alpha_1$ & $N$ & $\bar N$ &1&1 \\
\hline
$V^\alpha_2 $ & 1 & $ N$ &$\bar N$&1 \\
\hline
$U^\alpha_3$ & 1&1& $N$ & $\bar N$\\
\hline
$V^\alpha_4 $ & $\bar N$ & 1  &1& $N$ \\
\hline

\end{tabular}
\end{center}

All the above information can also be read off from the quiver diagram
\begin{center}
\includegraphics[width=.35\textwidth]{boxDiag.eps}
\end{center}

A candidate for the $A$ transformation is
\be
A: (U_1,V_2,U_3, V_4)\mapsto (U_3,V_4,U_1,V_2).
\ee

We now look for discrete transformations of the form
\be
(U_1,V_2,U_3, V_4)\mapsto (u_1U_1,v_2 V_2,u_3 U_3,v_4 V_4).
\ee
Invariance of the superpotential implies that
\be
u_1v_2u_3 v_4=1.
\ee
Anomaly cancelation implies that
\be
(u_1 v_2)^{2N}=(v_2u_3)^{2N}=(u_3 v_4)^{2N}=(v_4 u_1)^{2N}=1.
\ee

We find the following solutions for $B$ and $C$, with the condition
that $\omega^{2N}=1$:

\begin{eqnarray}
A&:& (U_1,V_2,U_3, V_4)\mapsto (U_3,V_4,U_1,V_2), \\ B&:&
(U_1,V_2,U_3, V_4)\mapsto (\omega U_1,V_2,U_3, \omega^{-1} V_4), \\
C&:& (U_1,V_2,U_3, V_4)\mapsto (\omega U_1,\omega V_2,\omega^{-1} U_3,
\omega^{-1} V_4)
\end{eqnarray}
\subsubsection{$\mathbb{Z}_4$ orbifold of the conifold }
Let us consider the quiver diagram corresponding to $Y^{4,0}$, which will
capture all of the features of the general case.

\begin{center}
\includegraphics[width=.75\textwidth]{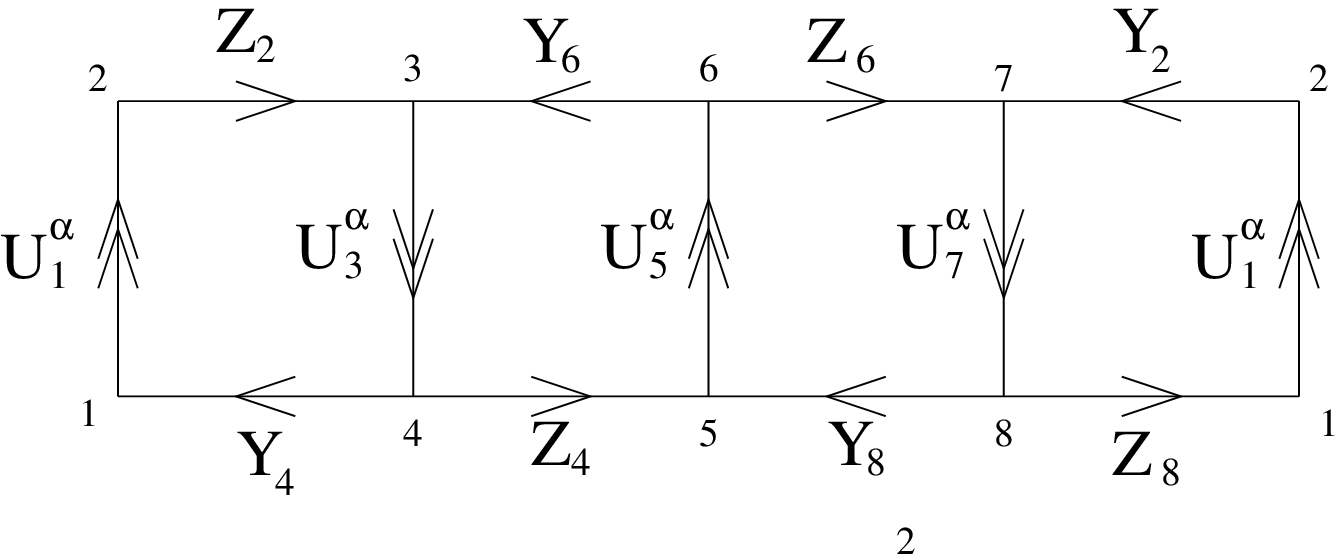}
\end{center}

Consider the $A$ element generated by
\be
A:
\begin{array}{ccc}
(U_1,U_3,U_5,U_7)& \mapsto &(U_7, U_1, U_3, U_5),  \\
(Z_2,Z_4,Z_6,Z_8) &\mapsto &(Z_8,Z_2,Z_4,Z_6 ),  \\  (Y_2,Y_4,Y_6,Y_8)
&\mapsto& (Y_8,Y_2,Y_4,Y_6 ).
\end{array}
\ee
Note that the most efficient way to represent this transformation is
by action on all kinds of fields separately.

We can also think of the $A$ transformation as a permutation on the
nodes:
\be
A: (1,3,5,7) \mapsto (7, 1,3, 5), \quad  (2,4,6,8)\mapsto (8,2,4,6).
\ee
The superpotential for this theory is
\be
W \sim \epsilon_{\alpha \beta} U^{\alpha}_1 Z_2 U_3^\beta Y_4
-\epsilon_{\alpha \beta} U^{\alpha}_3Z_4U_5^\beta Y_6  +
\epsilon_{\alpha \beta} U^{\alpha}_5Z_6U_7^\beta Y_8 -\epsilon_{\alpha
\beta} U^{\alpha}_7Z_8U_1^\beta Y_2.
\ee

Invariance of the superpotential implies that
\be
u_1 z_2 u_3 y_4=1, \quad  u_3 z_4 u_5 y_6=1, \quad  u_5 z_6 u_7 y_8=1,
\quad   u_7z_8u_1 y_2=1.
\ee

with an $\omega$ such that $\omega^{4N}=1$ we find that the remaining
transformations are:

\be
B:
\begin{array}{ccc}
(U_1,U_3,U_5,U_7)& \mapsto &(U_1, \omega U_3, \omega^2 U_5,\omega^3 U_7),  \\
(Z_2,Z_4,Z_6,Z_8) &\mapsto & \omega^{-3} (\omega^3 Z_2,\omega^2 Z_4,\omega Z_6, Z_8),
\\  (Y_2,Y_4,Y_6,Y_8) &\mapsto &  \omega^{-3}(\omega^3 Y_2,\omega^2 Y_4,\omega
Y_6,Y_8).
\end{array}
\ee

\be
C:
\begin{array}{ccc}
(U_1,U_3,U_5,U_7)& \mapsto &(\omega U_1,\omega U_3, \omega U_5,\omega U_7),  \\
(Z_2,Z_4,Z_6,Z_8) &\mapsto
&(\omega^{-1}Z_2,\omega^{-1}Z_4,\omega^{-1}Z_6,\omega^{-1}Z_8 ),  \\
(Y_2,Y_4,Y_6,Y_8) &\mapsto& (\omega^{-1}Y_2,\omega^{-1}Y_4,\omega^{-1} Y_6, \omega^{-1}Y_8).
\end{array}
\ee

Note that $B$ and $C$ are given as transformation acting on the three
sets of fields that we have, namely, $U, Y$ and $Z$.

Now we will check explicitly that the above transformations
satisfy the Heisenberg group laws.  First, let us note that we
will have to use the center of the gauge group to complete the
transformations.  The matrix representations of the above
transformations are block diagonal in the $U$ $Y$ and $Z$ fields.
One finds that these matrices obey
\be
AB=BAC\times
\begin{pmatrix}
\omega^4 & 0& 0& 0& 0& 0 \\
0    & {\mathbb I}_{3\times3} & 0 &0 &0 &0 \\
0&0& \omega^{-4} &0&0&0 \\
0&0&0&{\mathbb I}_{3\times3}&0&0\\
0&0&0&0&\omega^{-4}&0\\
0&0&0&0&0&{\mathbb I}_{3\times3}
\end{pmatrix}.
\ee
The last matrix on the right hand side is a member of the center of
the gauge group.  First note that because $\omega$ is a $4N^{\mbox{th}}$ root
of unity, that $\omega^4$ is an $N^{\mbox{th}}$ root of unity.  Therefore
to each gauge group we associate a $\omega^{4n_i}$, and rephase the
fields according to whether they are fundamental or antifundamental.
The center of the gauge group corresponding to the
above matrix is then given by
\begin{center}
\includegraphics[width=.75\textwidth]{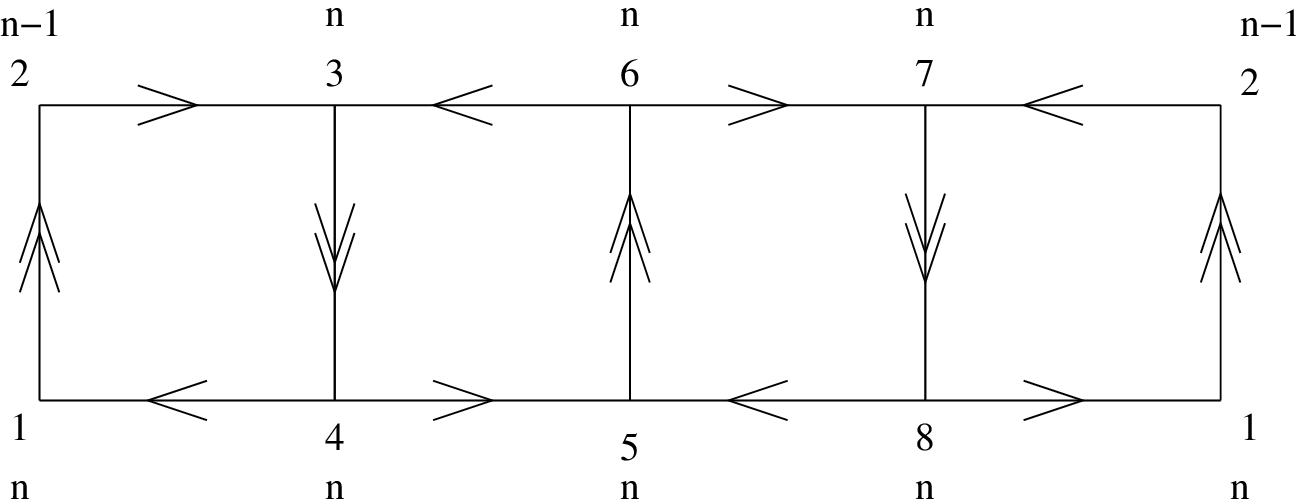}
\end{center}
for any integer $n$, and as before the $N^{\rm th}$ root of unity
associated with the center of the gauge group is $\gamma=\omega^4$.
The clock symmetry can shift this action to any of the other sets
of fields, acting on $(1,2),(3,4),(5,6)$ and $(7,8)$ fields
independently.  We also see that any of these operations
applied 4 times is gauge equivalent to the identity.  We therefore
find that
\be
A^{4}=B^{4}=C^{4}=1, \qquad AB=BAC.
\ee

\subsubsection{$\mathbb{Z}_p$ orbifold, the general case}
The procedure of the last subsection is easily lifted to the general $Y^{(p,0)}$ case.
The transformations are
\be
A:
\begin{array}{ccc}
(1,3,5,\cdots 2p-3, 2p-1)&\mapsto& (2p-1,1,3,\cdots,2p-5,2p-3),\\
(2,4,6\cdots, 2p-2,2p)   &\mapsto& (2p,2,4,\cdots,2p-4,2p-2).
\end{array}
\ee

\be
B:
\begin{array}{ccc}
(U_1,U_3,\cdots,U_{2p-1})& \mapsto &(U_1, \omega U_3, \omega^2 U_5,\cdots,\omega^{p-1} U_{2p-1}),  \\
(Z_2,Z_4,\cdots,Z_{2p}) &\mapsto &  (1 Z_2,\omega^{-1} Z_4,\cdots, \omega^{-(p-1)}Z_{2p}),\\
(Y_2,Y_4,\cdots,Y_{2p}) &\mapsto &  (1 Y_2,\omega^{-1} Y_4,\cdots, \omega^{-(p-1)}Y_{2p}).
\end{array}
\ee

\be
C:
\begin{array}{ccc}
(U_1,U_3,\cdots,U_{2p-1}) &\mapsto &(\omega U_1,\omega U_3, \cdots ,\omega U_{2p-1}),  \\
(Z_2,Z_4,\cdots,Z_{2p}) &\mapsto &(\omega^{-1}Z_2,\omega^{-1}Z_4,\cdots,\omega^{-1}Z_{2p} ),  \\
(Y_2,Y_4,\cdots,Y_{2p}) &\mapsto &(\omega^{-1}Y_2,\omega^{-1}Y_4,\cdots,\omega^{-1}Y_{2p} )
\end{array}
\ee
\be
\omega^{pN}=1.
\ee
These operations satisfy
\be
A^{p}=B^{p}=C^{p}=1, \qquad AB=BAC
\ee
up to the center of the gauge group.  The element of the center of the
gauge group discussed earlier is easily generalized:  promote
$4\rightarrow p$ and all new gauge groups are assigned
$pn$.

\subsection{Orbifolds of $Y^{p,q}$}
Another very interesting class of gauge theories are the quiver gauge theories obtained
as the gauge theory dual of string theory on $AdS_5\times Y^{p,q}$ with 5-form flux.
A very complete discussion of $Y^{p,q}$ spaces is presented in \cite{sasgeom}. The field theory
aspects are presented in \cite{ms,sequiver,friends}.

We now display some general results for orbifolds of $Y^{p,q}$, specifically
those that have $p$ and $q$ not relatively prime (this has the effect of
changing the periodicity of the $\beta$ angle, and so is an abelian orbifold
in this angle).  We will call the greatest common divisor of $p$ and $q$
$GCD(p,q)=s$.  We will follow the notation introduced in \cite{cells}, where an arbitrary
$Y^{p,q}$ quiver gauge theory is constructed as a sequence of two primitive
elements. The quiver diagram that we associate with this
geometry is that of $Y^{p/s,q/s}$ repeated $s$ times:
\be
\underbrace{\underbrace{(\sigma \tilde{\sigma}\tau \cdots\cdots\cdots\cdots\cdots\cdots)}
_{((p-q)/a)\;\tau{\rm-type},\; (q/a)\; \sigma{\rm-type}}(\cdots)(\cdots)\cdots}_{s{\rm -times}}.
\ee
where $\sigma$ and $\tau$ are the unit cells
\begin{figure}[h]
\centering
\includegraphics[width=.5\textwidth]{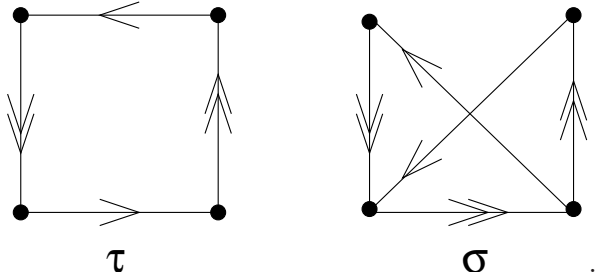}.
\caption{The unit cells $\tau$ and $\sigma$.}
\end{figure}

For details of constructing the quivers gauge theories for $Y^{p,q}$
we refer the reader to \cite{sequiver, cells}.
Because the ends are identified one may rearrange the above diagrams into polygon diagrams which
have $2p$ sides, and a number of internal line, which depends on the particular value of $q$.  The
$A$ transformation is then cyclicly mapping the primitive $Y^{p/s,q/s}$ cells into each other, and
so is a $\zet_s$ symmetry.

We now go about finding the rephasing symmetries associated with such diagrams.  First, we will
use the fact that the superpotential terms allow us to eliminate the internal lines, and so
we may discuss only the phases of the sides of the polygon.  For brevity, we will always label
the fields on the outside with $U_i$ and the subscript will denote the node that the arrow
is pointing away from.  The internal lines will be labeled as $Z_i$ for terms entering in cubic
superpotential terms and $Y_i$ for terms entering in quartic terms, with the $i$ now denoting
the node that the arrow is pointing towards.  The node labeled $1$ will always precede 3 double
arrows (which we are guaranteed to have because we assume at least one $\tau$ type cell).
With this labeling, we may write down a set of
rules that determines the anomaly cancelation condition for an arbitrary node in the diagram.
The rule is based purely on the previous 2 and following two fields on the edge of the polygon.
The previous two fields are either two double lines $(D,D)$, a single
line and then a double
line $(S,D)$
or a double and then a single line $(D,S)$.  The fields coming after the node fall into the same
categories.  From this information, the anomaly condition at the $m^{\rm th}$ node
\begin{center}
\includegraphics[width=.5\textwidth]{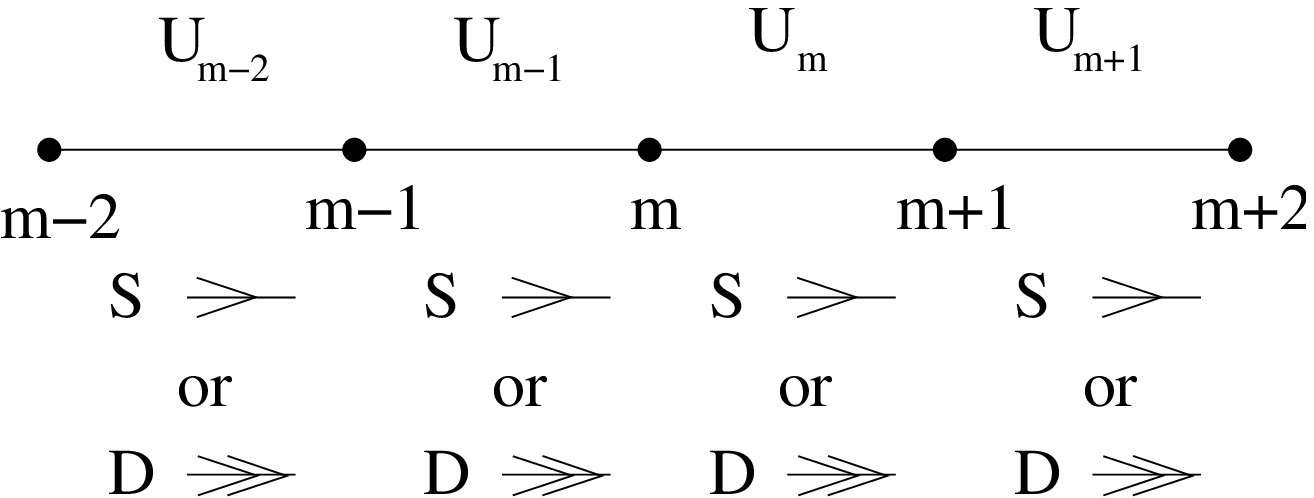}.
\end{center}
is given by
\be
\label{genanom}
\left(\begin{array}{ccc}
\left\{
\begin{array}{ccc}
(D,D)& \mapsto & \frac{1}{u_{m-2}} \\
(S,D)& \mapsto & \frac{1}{u_{m-3}u_{m-2}} \\
(D,S)& \mapsto & 1
\end{array}
\right\}
& \times u_{m-1} u_m \times &
\left\{
\begin{array}{ccc}
(D,D)& \mapsto & \frac{1}{u_{m+2}} \\
(S,D)& \mapsto & 1 \\
(D,S)& \mapsto & \frac{1}{u_{m+1}u_{m+2}}
\end{array}
\right\}
\end{array}\right)^N=1
\ee
In addition to these, we want that the scalings
are naturally associated with a member of the center
of the gauge group, and so we require that
\be
\label{product}
\prod u_i=1.
\ee

Now, to fix the $B$ transformation, we set $a_1=1,a_2=\omega,a_3=\omega^2$.
We take the anomaly conditions from node $3$ and forward to $2p-2$, and
require that they are trivially satisfied (we mean that the quantity
appearing in (\ref{genanom}) is $1$ before raising to the $N^{\rm th}$
power).  Taking into account enough equations, $a_1, a_2$ and $a_3$
will determine $a_4$, and so on up to $a_{2p-1}$.  The quantity
$a_{2p-1}$ is then determined by (\ref{product}).  The remaining 4 equations are
``boundary'' terms, involved in consistently ``gluing'' the ends
together.  They will give conditions of the form $\omega^{a_i \times N}=1$
for $i=1,2,2p,2p-1$.  The $a_i$ should have common divisor of $s$ where
again $GCD(p,q)=s$.  This gives us the requirement that $\omega^{sN}=1$.  Rasing
this $B$ operation to the $s^{\rm th}$ power automatically is in the center of
the gauge group.

To determine the $C$ operation, first note that the $A$ operation naturally
associates sets of fields that transform into each other.  In particular
$U_i$ is associated with $U_{i+2p/a}$, which is associated with $U_{i+4p/a}$
etcetera.  From this association, we may associate their scalings.  They should
satisfy $u_{i+4p/s}=\omega^{k_i}u_{i+2p/s}=\omega^{2k_i}u_{i}$.  We may read off $k_1,k_2$
and $k_3$ from the first set of fields.  We now repeat the procedure above,
only we seed with $a_1=\omega^{k_1},a_2=\omega^{k_2},a_3={\omega^{k_3}}$.  Again, the
fact that $C$ is an $s^{\rm th}$ root of a member of the center of the gauge
group is true by construction.

To be more concrete, we work an example.  Namely, we will work out $A$, $B$ and $C$
transformations for the quiver of $Y^{6,3}$:
\begin{center}
\includegraphics[width=.5\textwidth]{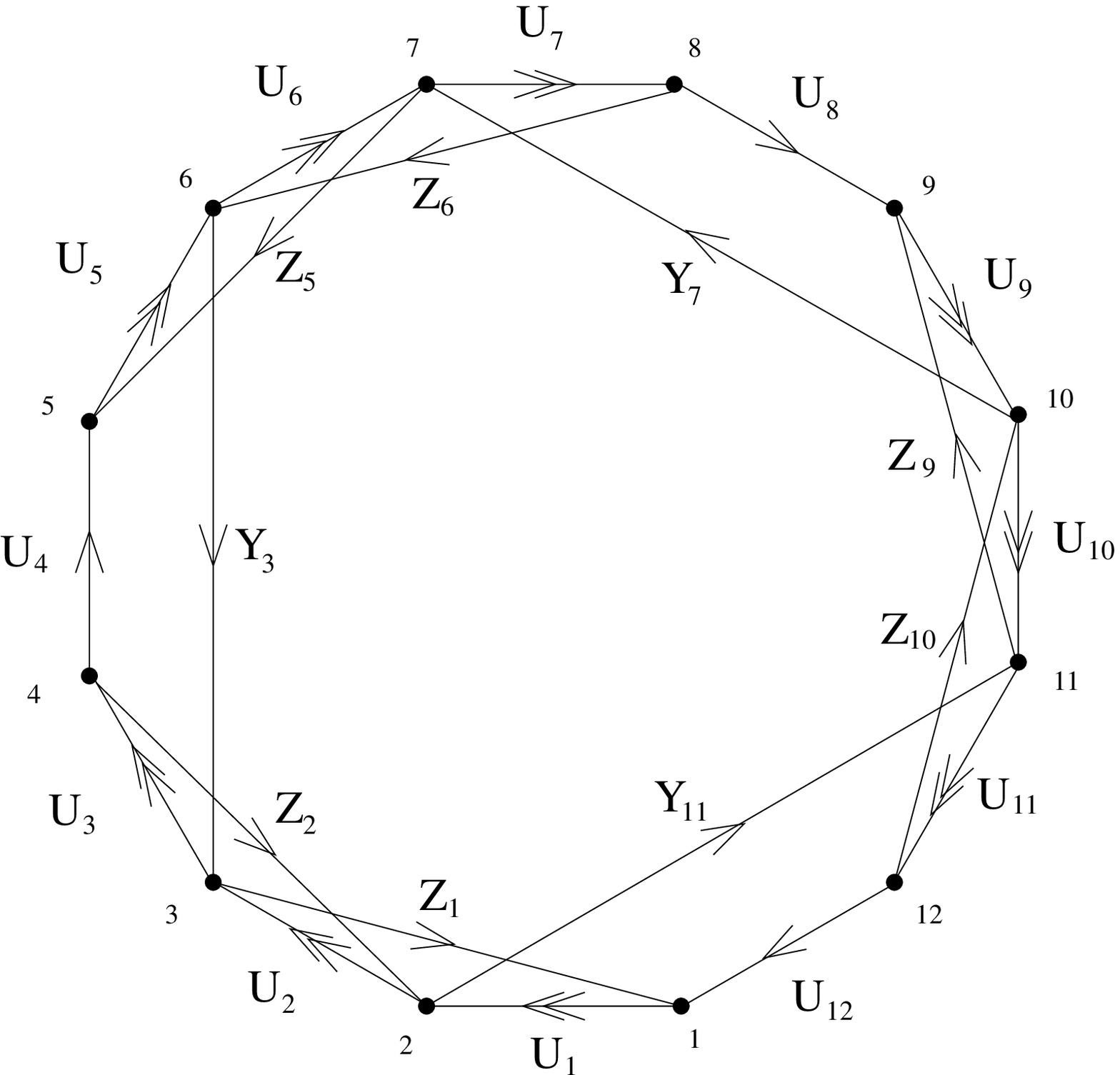}.
\end{center}
First, the $A$ transformation is easy to pick out as being
\be
A:
\begin{array}{ccc}
(1,5,9)&\mapsto& (9,1,5),\\
(2,6,10)&\mapsto& (10,2,6)\\
(3,7,11)&\mapsto& (11,3,7)\\
(4,8,12)&\mapsto& (12,4,8).
\end{array}
\ee
Next, the anomaly cancelation conditions are
\bea
\left(\frac{u_{12} u_1}{u_2}\right)^N &=&1, \quad
\left(\frac{u_1 u_2}{u_{11} u_{12} u_{3}}\right)^N =1, \nn \\
\left(\frac{u_2 u_3}{u_1 u_4 u_5}\right)^N &=&1, \quad
\left(\frac{u_3 u_4}{u_2}\right)^N =1, \nn\\
\left(\frac{u_4 u_5}{u_6}\right)^N &=&1, \quad
\left(\frac{u_5 u_6}{u_3 u_4 u_7}\right)^N =1, \label{anomCond} \\
\left(\frac{u_6 u_7}{u_5 u_8 u_9}\right)^N &=&1, \quad
\left(\frac{u_7 u_8}{u_6}\right)^N =1, \nn\\
\left(\frac{u_8 u_9}{u_{10}}\right)^N &=&1, \quad
\left(\frac{u_9 u_{10}}{u_7 u_8 u_{11}}\right)^N =1, \nn\\
\left(\frac{u_{10} u_{11}}{u_{9}u_{12}u_{1}}\right)^N &=&1,\quad
\left(\frac{u_{11} u_{12}}{u_{10}}\right)^N =1\nn.
\eea
We begin solving the equations as prescribed above, solving the $2^{\rm nd}$
through $5^{\rm th}$ line trivially using the prescription defined above.  The
solution for the $B$ transformation is
\be
B: U_{i}\mapsto u_i U_{i},\; Z_{i}\mapsto z_i Z_{i},\; Y_{i}\mapsto y_i Y_{i}
\ee
with
\be
\begin{array}{c c c}
u_1=1 & u_5= \omega^4 & u_9=\omega^8 \\
u_2=\omega & u_6= \omega^3 & u_{10}=\omega^5 \\
u_3=\omega^2 & u_7= \omega^6 & u_{11}=\omega^{10} \\
u_4=\omega^{-1} & u_8= \omega^{-3} & u_{12}=\omega^{-35} \\
\end{array}.
\ee
and
\be
\begin{array}{c c c}
z_1=\omega^{-1} & z_5= \omega^{-7} & z_9=\omega^{-13} \\
z_2=\omega^{-3} & z_6= \omega^{-9} & z_{10}=\omega^{-15} \\
y_3=\omega^{-5} & y_7= \omega^{-11} & y_{11}=\omega^{25}
\end{array}.
\ee
The set of $z$ and $y$ variables are simply read off from the
superpotential constraints.  We find $k_1=4,k_2=2,k_3=4$.
Repeating the above procedure, we find that
\be
C: U_{i}\mapsto u_i U_{i},\; Z_{i}\mapsto z_i Z_{i},\; Y_{i}\mapsto y_i Y_{i}
\ee
with
\be
\begin{array}{c c c}
u_1=\omega^4 & u_5= \omega^4 & u_9=\omega^4 \\
u_2=\omega^2 & u_6= \omega^2 & u_{10}=\omega^2 \\
u_3=\omega^4 & u_7= \omega^4 & u_{11}=\omega^{4} \\
u_4=\omega^{-2} & u_8= \omega^{-2} & u_{12}=\omega^{-26} \\
\end{array}.
\ee
and
\be
\begin{array}{c c c}
z_1=\omega^{-6} & z_5= \omega^{-6} & z_9=\omega^{-6} \\
z_2=\omega^{-6} & z_6= \omega^{-6} & z_{10}=\omega^{-6} \\
y_3=\omega^{-6} & y_7= \omega^{-6} & y_{11}=\omega^{18}
\end{array}.
\ee
The final requirements from the first and last lines of (\ref{anomCond})
are satisfied for $\omega^{3N}=1$ (actually, one could have the less restrictive
$\omega^{6N}=1$, however, we wish to find objects such that they are one when
raised to the $3^{\rm rd}$ power, so that the order matches that of $A$).
One may check that $A$ and $C$ commute up to the center of the gauge group
generated by
\begin{center}
\includegraphics[width=.5\textwidth]{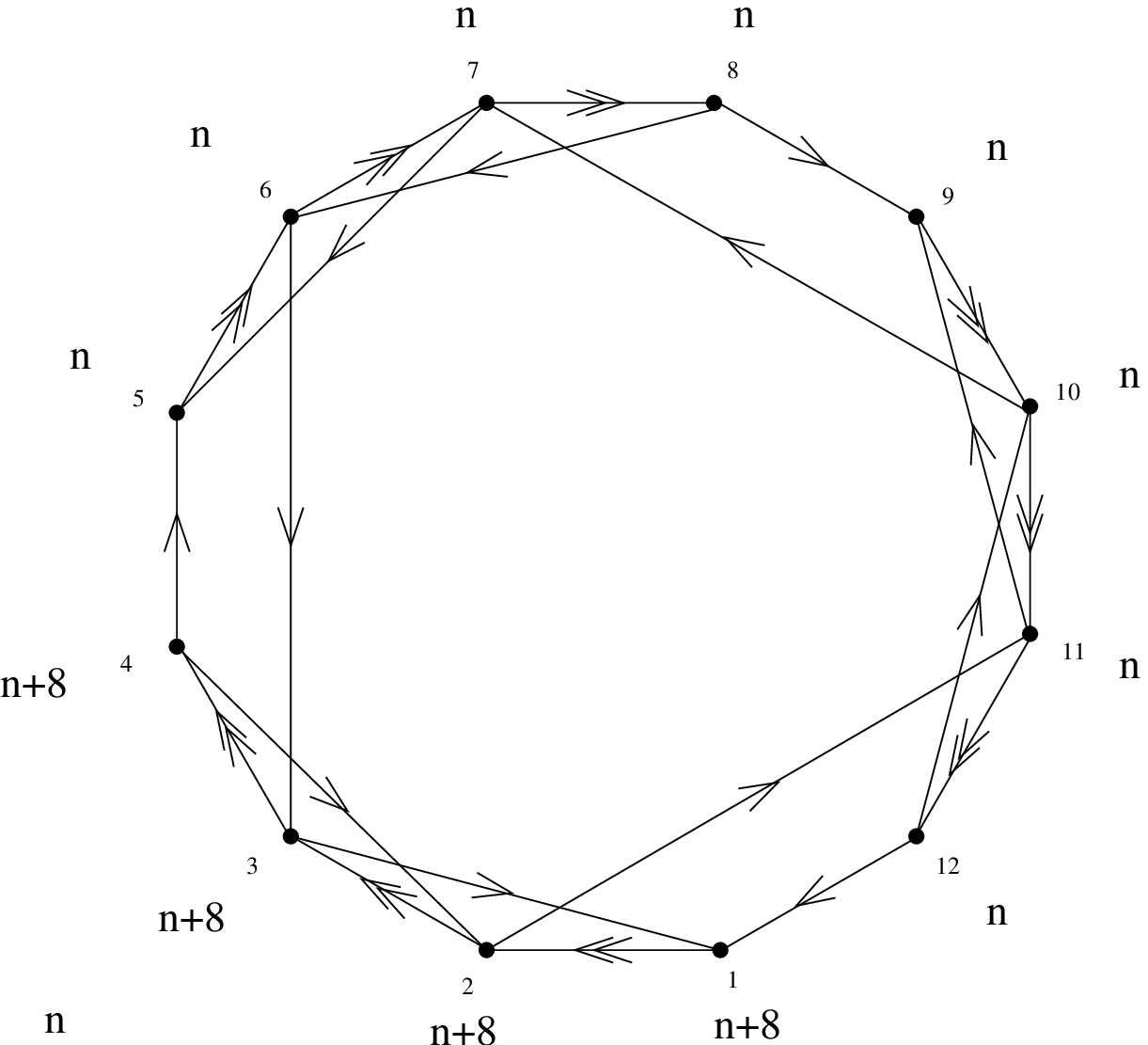}.
\end{center}
One may also check that
\be
AB=BAC
\ee
up to the center of the gauge group associated with
\begin{center}
\includegraphics[width=.5\textwidth]{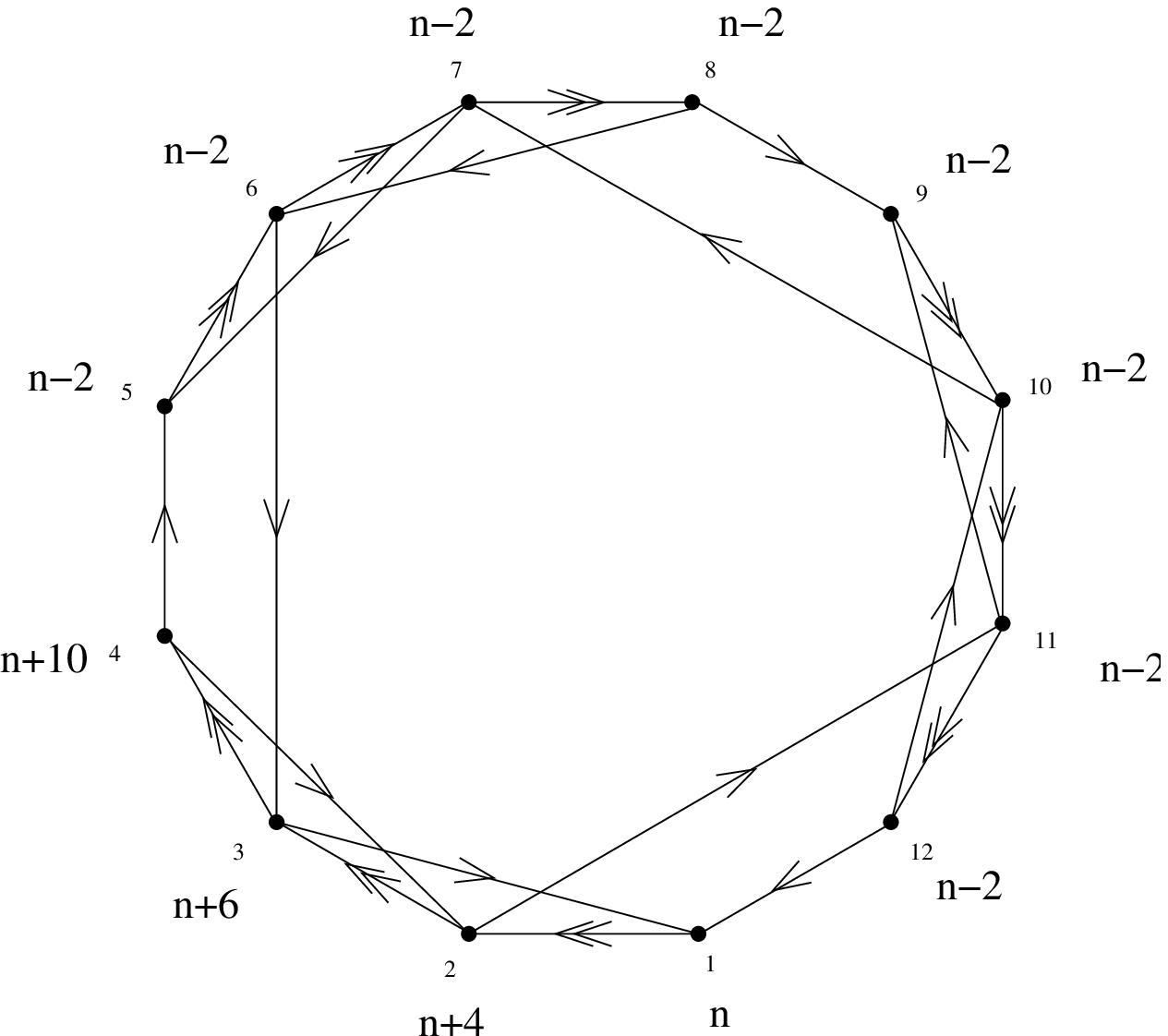}.
\end{center}
Again, in these diagrams one takes the center of the gauge group and
rotates by $\omega^{3n_i}$ and then the charges of the fields under the
$SU(N)$ determines the rephasing of the fields.  We also find that
the above operations satisfy $A^3=B^3=C^3=1$ up to an element in the center of the gauge group.

\section{D-brane interpretation}
As suggested in \cite{grw} the field theory results can be matched
with states in the dual string theory. Namely, one  can think of the
above operators $A,B$ and $C$ as the operators counting the number of
fundamental strings  and D-strings wrapped around a 1-cycle  and $C$
as the number of D3 branes wrapped on that cycle. Alternatively, one
can  think of $A$ and $B$ as the operators counting the number of NS5
branes and D5 branes wrapping  a 3-cycles and $C$  the operator
counting the number of D3 branes wrapping a 1-cycle.

Another powerful point of view discussed in \cite{grw} and
more recently in \cite{dima} identifies the $A$ and $B$ symmetries
with Wilson and t' Hooft loops. The idea is that a wrapped D3 branes
supports a $U(1)$ gauge field in its worldvolume. This gauge field
basically carries fundamental and D string numbers in the guise of
eigenvalues of Wilson and t' Hooft loops. This point of view is
powerful because, at low energies, it reduces the problem to a $U(1)$ gauge field in a
nontrivial worldvolume, much about this situation has been discussed
in \cite{freed}. One expects that topological arguments are then
independent of the low energy limit.

Returning to the string theory interpretation.
The $A$ symmetry, being perturbative, is naturally identified with
perturbative states in the string theory, i. e.,  the fundamental
string. An important piece of evidence in identifying the string
states comes from the fact that  Heisenberg groups admit an  action of $SL(2,\mathbb{Z})$ which we
interpret as a symmetry of type IIB string theory. Under an
$SL(2,\mathbb{Z})$ transformation given by the matrix

\be
\label{sl2}
M=\left(
\begin{array}{cc}
a&b\\ c&d\\
\end{array}
\right),
\ee
the operators transform as
\be
A\mapsto A^a B^b, \quad B\mapsto A^c B^d, \quad C\mapsto C.
\ee

One important check in the identification of the string theory states
is that the number of states coming from wrapped branes and wrapped
fundamental strings is given by
the existence of the corresponding cycles in the dual string theory. For example, for orbifolds of
$\mathbb{C}^3$ the relevant string theory is $AdS_5\times
S^5/\mathbb{Z}_q$. The
calculation of the homology
groups is different from the one presented in the appendix of \cite{grw}. The main difficulty is that
the resulting space is not a Lens space due to the fact that the
orbifold action
is not compatible with the
action of the Hopf fibration $S^1\longrightarrow S^5 \longrightarrow
\mathbb{CP}^2$. To see this recall that the orbifold
action is inherited from the action on $\mathbb{C}^3$ given as
$(z_1,z_2,z_3)\mapsto (\omega z_1, \omega z_2, \omega^{-2}z_3)$ where $\omega$ is
a $q$-th root of unity. This action was used in section two. On the
other hand, the Lens space is define as the quotient of
$S^{5}$ by the action of $\mathbb{Z}_q: (z_1,z_2,z_3)\mapsto (\omega z_1,
\omega z_2, \omega z_3)$ which is, of course compatible with the Hopf
fibration. Note that only
for $q=3$ the gauge theory orbifold is
isomorphic to the orbifold in the definition  of Lens space.
Nevertheless, using the Leray spectral sequence one finds that, the relevant torsion
terms are given by
\bea
H_1(S^5/\mathbb{Z}_q)&=&\mathbb{Z}_q, \nonumber \\
H_3(S^5/\mathbb{Z}_q)&=&\mathbb{Z}_q.
\eea
This is in agreement with our discussion of orbifolds of ${\cal N}=4$ SYM theory in section two.

The calculation of homology groups in the case of the $T^{1,1}$ and, in
general for $Y^{p,q}$ spaces is less
straightforward. However, here too there are various approaches.
In one approach it is convenient to consider the cone over these spaces
which are toric varieties for which the relevant  properties are known. Alternatively \cite{ms}, one
attacks the calculation by viewing these spaces as $U(1)$ bundles over
some K\"ahler-Einstein base. The relevant results are compiled in
\cite{sasgeom,ms}. The most relevant statement follows from  Lerman \cite{lerman} and
states that the fundamental group of $Y^{p,q}$ manifolds is
\be
\pi_1(Y^{p,q})=\mathbb{Z}_h,
\ee
where $h$ is the highest common factor of $(p,q)$.  Further we use,  that the fundamental group is
isomorphic to the first homology group which coincides
with the third homology group.  Note that this is precisely the fact that was exploited in constructing
the $A$ transformation in section two.

\subsection*{A correction for $SL(2,\mathbb{Z})$ in the presence of flux?}
We see that the $SL(2,\mathbb{Z})$ action, as written above, is broken in the general case.
The reason is its incompatibility with the cyclic property of the group elements.
To understand the problem note that if we perform an $SL(2,\mathbb{Z})$  transformation given by the matrix
(\ref{sl2}) then the cyclic property of, say, $A$ is affected:
\bea
A^q&\mapsto& \underbrace{A^a B^b \ldots A^a B^b}_{q -{\rm times}} \nonumber \\
& = & A^{qa} B^{qb} C^{-\frac12 q(q+1) ab}.
\eea
Given that $A^q=B^q=C^q=1$ this is not a problem if $q$ is odd. Namely, for $q=2r+1$ odd,
we have that after the transformation $A^q\mapsto C^{q}{}^{(-(r+1) a
b)}$ which is
the unity transformation and therefore
cyclicity is preserved. However for $q=2r$ even we see
that $A^q \mapsto C^q{}^{(-\frac12 (2r+1)  ab)}$ which is not
necessarily  the identity transformation.  The failure, however, is
restricted to a square root of minus one.

One can attempt to remedy the situation by suggesting that the $SL(2, \mathbb{Z})$  action is
\be
M: \quad A\mapsto A^a B^b C^{ab/2} , \quad B\mapsto A^c B^dC^{cd/2} , \quad C\mapsto C.
\ee
Note, since $C$ commutes with $A$ and $B$ this is like simply
multiplying by some $c$-number.

\section{Conclusions}

In this note we have explicitly demonstrated the existence of finite Heisenberg groups
realized as discrete transformation in a large class of quiver gauge theories. We believe our
work has interest implications for our understanding of the D-branes.

Given the current state of string theory technology, understanding branes in
backgrounds with Ramond-Ramond fluxes is beyond our means. In this sense the AdS/CFT
provides a unique opportunity to study the nature of D-branes. In this
case, we established
indirectly that  in
backgrounds with nontrivial RR flux, if the space has torsion classes in the homology groups, the
operators whose iegenvalues determine the number of branes wrapping
the corresponding cycles satisfy a finite  Heisenberg group.

There are various issues that remain to be explored and we hope to
return to some of them in the near future.  First there is the question of
whether most quiver gauge theories admit such structure, in
particular if all  toric quivers admit such construction. As we
suspect a necessary condition is given by the existence of torsion in
the first and third homology classes of the horizon geometry. Another
interesting question, which we plan to explore is the relationship between
the finite Heisenberg groups  and Seiberg duality in the quiver. It would
also be interested to consider  nonconformal situations. Naively,
given the conditions we impose in the construction of the operators
$A,B$ and $C$, one should  not expect such symmetry to survive the
noncorformal limit. However, if this is a general property of D-branes
there should an action of a finite Heisenberg group.
More generally one wonders if noncommutativity is a property of
D-branes in backgrounds with RR flux how could one  approach a similar
calculation for IIA string theory.

\section*{Acknowledgments}
We are grateful to J. Gauntlett and Y. Tachikawa for comments and correspondence. We
are thankful to G. Moore  for posing the question of Heisenberg
groups in quiver theories in the course of  the workshop
``Mathematical Structures in String Theory''  at the KITP.  We are
particularly grateful to D. Belov for various comments and
clarifications. This work
is  partially supported by Department of Energy under grant
DE-FG02-95ER40899 to the University of Michigan and by the National
Science Foundation under Grant No. PHY99-07949 to the KITP.

\end{document}